\documentclass[showpacs,prl,onecolumn,aps,superscriptaddress,preprintnumbers,nofootinbib]{revtex4}
\usepackage[T1]{fontenc}
\usepackage[latin9]{inputenc}
\usepackage{amsmath,amssymb}
\usepackage{epsfig}
\usepackage{graphicx}
\usepackage{amsmath}
\usepackage{amsfonts}
\def\slashchar#1{\setbox0=\hbox{$#1$}     		
   \dimen0=\wd0                                 	
   \setbox1=\hbox{/} \dimen1=\wd1               	
   \ifdim\dimen0>\dimen1                        	
      \rlap{\hbox to \dimen0{\hfil/\hfil}}      	
      #1                                        	
   \else                                        	
      \rlap{\hbox to \dimen1{\hfil$#1$\hfil}}   	
      /                                         	
   \fi}

\renewcommand{\vec}{\boldsymbol}
\newcommand{\beq}{\begin{equation}}
\newcommand{\eeq}{\end{equation}}
\newcommand{\bea}{\begin{eqnarray}}
\newcommand{\eea}{\end{eqnarray}}
\newcommand{\ba}{\begin{array}}
\newcommand{\ea}{\end{array}}

\def\eq#1{{Eq.~(\ref{#1})}}
\def\fig#1{{Fig.~\ref{#1}}}
\newcommand{\bas}{\bar{\alpha}_S}

\newcommand{\nn}{\nonumber}

\newcommand{\h}{\frac{1}{2}}

\newcommand{\Lb}{\left(}
\newcommand{\Rb}{\right)}
\def\pom{{I\!\!P}}
\def\reg{{I\!\!R}}
\begin{document}
\title{CGC/saturation approach: secondary Reggeons and $\rho={ \rm Re/Im}$  dependence on energy}
\author{E. ~Gotsman}
\email{gotsman@post.tau.ac.il}
\affiliation{Department of Particle Physics, School of Physics and Astronomy,
Raymond and Beverly Sackler
 Faculty of Exact Science, Tel Aviv University, Tel Aviv, 69978, Israel}
 \author{E.~ Levin}
 \email{leving@post.tau.ac.il, eugeny.levin@usm.cl}
 \affiliation{Department of Particle Physics, School of Physics and Astronomy,
Raymond and Beverly Sackler
 Faculty of Exact Science, Tel Aviv University, Tel Aviv, 69978, Israel}
 \affiliation{Departemento de F\'isica, Universidad T\'ecnica Federico
 Santa Mar\'ia, and Centro Cient\'ifico-\\
Tecnol\'ogico de Valpara\'iso, Avda. Espana 1680, Casilla 110-V,
 Valpara\'iso, Chile}  
 \author{I. Potashnikova}
 \email{irina.potashnikova@usm.cl}
  \affiliation{Departemento de F\'isica, Universidad T\'ecnica Federico
 Santa Mar\'ia, and Centro Cient\'ifico-\\
Tecnol\'ogico de Valpara\'iso, Avda. Espana 1680, Casilla 110-V,
 Valpara\'iso, Chile}  
\date{\today}




\begin{abstract} In this letter we demonstrate that a  model  
based
 on the CGC/saturation approach, successfully describes  soft interaction 
collisions  for the wide range of $W=\sqrt{s} = 30 \,GeV 
 \div\, 13 \,TeV$ including   the new TOTEM data at 13 TeV.
 We have now incorporated the secondary Reggeons in our approach this
 enables us to describe the complete set of the soft data, including
 the energy behaviour of $\rho $  the ratio of the real to imaginary
parts of the elastic scattering amplitude. We argue that
 it is premature to claim that an  odderon contribution is necessary, but
 estimate its possible strength as $1 mb$ to the real part
 of the amplitude  at W = 13 \,TeV. We show that the odderon contribution 
depends on the value of energy leading to ${\rm Re A\Lb s, t=0\Rb} $= 8 mb  at
 W=21.2 \,GeV. Bearing this in mind we do not believe that $\rho$ at high
 energies is the appropriate observable for detecting the odderon 
contribution.
The successful  description of the soft data in the wide
 range of energy  strengthens
 our   claim that the CGC/saturation approach is  the  only viable 
candidate for an  effective theory for high energy QCD.

\end{abstract}

\maketitle


{\it Introduction:}\\

In our recent papers\cite{GLP0,GLP} we have constructed a model,
 which allows us
 to discuss
  soft and  hard processes on the same footing. This model is based on
 the CGC/saturation approach(see Ref.\cite{KOLEB} for a review) and on
 our previous attempts to build such a model
\cite{GLMNI,GLM2CH,GLMINCL,GLMCOR,GLMSP,GLMACOR,GOLE,GOLELAST}.

The model, that we proposed in Refs.\cite{GLP0,GLP},
   successfully
describes the 
DIS data from HERA, the total, inelastic, elastic and
 diffractive cross sections, the $t$-dependence of
 these cross sections, as well as the inclusive production and rapidity 
and angular
 correlations in a wide range of high energies up to 13 TeV. 
 In this letter we include in our formalism the contribution 
of the secondary Reggeons that allows us to expand the region 
of energies down to $W = 50 \,GeV$. The main motivation for this
 is our need to describe the energy behaviour of the ratio $\rho
 = {\rm Re}/{\rm Im}$  of the scattering amplitude, to which the
 secondary reggeons make a significant contribution.  The new  TOTEM
 data \cite{TOTEMRHO} shows that the value of  $\rho = 0.1 \pm 0.01
 ( 0.09 \pm 0.01)$ is  lower than expected, and this ignited the hot 
discussion
 about the possible contribution of the odderon\cite{KMRO,BJRS,TT,MN,BLM,SS,
KMRO1,KMRO2}.  The odderon contribution to the negative signature appears
 naturally
  in perturbative QCD, with an intercept which is equal to 
zero\cite{BLV,KS}. In other words, we expect that the odderon will 
lead to a constant  real part of the scattering amplitude  at high
 energy, while the other contributions from the pomeron and the secondary
 reggeons  result in a decreasing real part. 
 Consequently, it is apt to elucidate the odderon
 contribution by     investigating $\rho$ at high energies.

 The main result of this letter is (i) that $\rho \approx 0.1$ at
 $W = 13\,TeV$  appears naturally in our approach without having to
assume an odderon contribution; and (ii) the value of its possible  
contribution is about $1 \,mb$ in the real part of the scattering
 amplitude, independent of energy. The small  and decreasing value of 
$\rho$ at 
high energies  stems from the CGC motivated amount of  shadowing
 (screening ) corrections and   furnishes   strong support
 for our approach.


Our model  incorporates two
 ingredients: the achievements of the CGC/saturation approach, and the 
pure  
phenomenological
 treatment of the long distance non-perturbative  physics, necessary, due 
to the lack of
  theoretical understanding of the confinement of quark and gluons.

  
     {\it The model: theoretical background for BFKL Pomerons and their
 interaction.}
 

  In Ref.\cite{AKLL} it  is shown that at high energy,  but in 
 the energy range:
 \beq \label{MOD1}
Y \,\leq\,\frac{2}{\Delta_{\mbox{\tiny BFKL}}}\,\ln\Lb
 \frac{1}{\Delta^2_{\mbox
{\tiny BFKL}}}\Rb
\eeq 
    Pomerons and their interactions, determine
 the scattering amplitude. In other words, it is shown that in
 this energy range the CGC approach and the BFKL Pomeron calculus
 are equivalent. In this paper   it is also been shown  that we can use
 the
  MPSI approximation\cite{MPSI}, to find the
 dressed Green function of the BFKL Pomeron ($G^{\mbox{\tiny dressed}}
\Lb r,
 Y - Y_0, \vec{b}\Rb$). This Green function exhibits a  geometric 
scaling behavior,  being function of one variable $ \tau = r^2 Q^2_S\Lb Y, 
b\Rb$, where $r$ denotes the size of the colorless dipole and $Q_s$  the 
saturation momentum; and it can be found using the solution to the 
non-linear 
Balitsky-Kovchegov
 equation \cite{BK}.

\begin{figure}[ht]
    \centering
  \leavevmode
      \includegraphics[width=14cm]{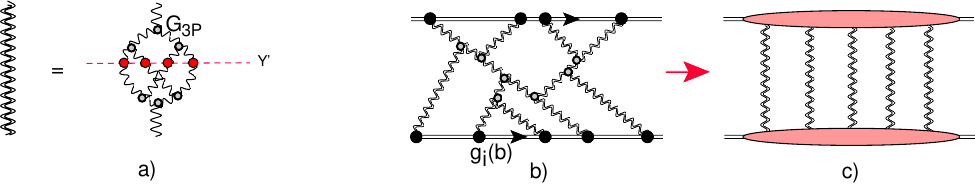}
   \caption{\protect\fig{amp}-a shows the set of  diagrams in the
 BFKL Pomeron calculus that produce the resulting (dressed) Green
 function of the Pomeron in the framework of high energy QCD. The red blobs
 denote the amplitude for the dipole-dipole interaction at low energy.
 In \protect \fig{amp}-b the net diagrams,    which   include
 the interaction of the BFKL Pomerons with colliding hadrons, are shown.
 The sum of the diagrams after integration
 over positions of $G_{3 \pom}$ in rapidity, reduces to  
 \protect\fig{amp}-c. The wavy lines denote the 
 BFKL Pomerons, while the double wavy lines describe the dressed 
Pomerons.}
\label{amp}
   \end{figure}

 Based on the above results  we constructed  a 
model\cite{LEPP,GLMNI,GLM2CH,GLP0,GLP} which describes the
 soft interaction at high energy in the region given by \eq{MOD1}.
  In this equation
  $\Delta_{\mbox{\tiny BFKL}}$ denotes the intercept of the BFKL 
 Pomeron. It turns out that in our model $ \Delta_{\mbox{\tiny BFKL}}\,
\approx\,0.2 - 0.25$    leading to $Y_{max} = 20 - 30$, which covers
 all collider energies.  The procedure of  calculation
 of $G^{\mbox{\tiny dressed}}
\Lb r,
 Y - Y_0, \vec{b}\Rb$ is shown in \fig{amp}-a, which  illustrates
 the MPSI approximation. The `fan' Pomeron diagrams  in \fig{amp}-a 
 correspond to the solution of the  nonlinear BK equation. We found a
 numerical solution to the BK equation, and an analytical formula that 
describes this solution to within 5\% accuracy
 (see Refs.\cite{LEPP,GLMNI,GLP,GLP0}). Finally, 
 
 \beq \label{G}
G^{\mbox{\tiny dressed}}\Lb G_\pom\Lb z \Rb \Rb = a^2 (1
 - \exp\Lb -G_\pom\Lb z \Rb\Rb )  +
 2 a (1 - a)\frac{G_\pom\Lb z \Rb}{1 + G_\pom\Lb z \Rb} + (1 -
 a)^2 \widetilde{G}\Lb G_\pom\Lb z \Rb\Rb
 \eeq
with $\widetilde{G}\Lb T\Rb = 1 - \frac{1}{T} \exp\Lb \frac{1}{T}\Rb
 \Gamma\Lb 0, \frac{1}{T}\Rb$
where $\Gamma\Lb s, z\Rb$ is the upper incomplete gamma function
 (see Ref.\cite{RY} formula {\bf 8.35}). 
  $G_\pom\Lb z \Rb$ denotes  the BFKL Pomeron 
in the vicinity of the saturation scale: 
\beq \label{GPZ}
 G_\pom\Lb z \Rb \,\,=\,\,\phi_0 \Lb r^2 \,Q^2_s\Lb Y-Y_0,b\Rb\Rb^{1
 - \gamma_{cr}}.
\eeq
 
  $a = 0.65$ for the solution of the  BK equation. For the saturation 
momentum
 we use the following general  formula
   \beq \label{QS}
Q^2_s\Lb b, Y\Rb\,\,=\,\,Q^2_{0s}\Lb b, Y_0\Rb\,e^{\lambda \,(Y - Y_0)}
~~\mbox{where}~~
Q^2_{0s}\Lb b, Y_0\Rb\,\,=\,\, \Lb m^2\Rb^{1 - 1/\bar \gamma}\,\Lb S\Lb b,
 m\Rb\Rb^{1/\bar{\gamma}} \,\mbox{with}\,\,\,
S\Lb b,m\Rb= \frac{m^2}{2 \pi} e^{ - m b}
\eeq 
In \eq{G} - \eq{QS} we know from    leading order   perturbative QCD,
 that $\lambda = 4.8 \bas$ and  $\bar{\gamma} = 1 - \gamma_{cr}
 = 0.63$. $\phi_0$ and  $m$ are the phenomenological parameters which
 determine the value of the Pomeron Green function at $\tau =1$ (from
 \fig{amp}-a one can see that   
$\phi_0 \,\propto \,$ dipole-dipole amplitude) and the typical dimensional
 scale at $Y =Y_0$. In \eq{QS} we  specify the large impact parameter behavior
 which cannot be derived from CGC approach \cite{KOWI}.  The form
 of $b$ dependence is purely phenomenological, while the exponential 
decrease at
 large $b$ follows from the Froissart theorem \cite{FROI}.

  
   {\it The model: phenomenological input for the hadron structures.}
 
 However, we cannot  build the scattering amplitude without  a
 phenomenological input for  the structure of the scattering
 hadrons. For this, we use a two channel model, which allows us to 
calculate the
 diffractive production in the region of small masses.
   In our model, we replace the rich structure of the 
 diffractively produced states, by a single  state with the wave 
function 
$\psi_D$, a la Good-Walker \cite{GW}.
  The observed physical 
hadronic and diffractive states are written in the form 
\beq \label{MF1}
\psi_h\,=\,\alpha\,\Psi_1+\beta\,\Psi_2\,;\,\,\,\,\,\,\,\,\,\,
\psi_D\,=\,-\beta\,\Psi_1+\alpha \,\Psi_2;~~~~~~~~~
\mbox{where}~~~~~~~ \alpha^2+\beta^2\,=\,1;
\eeq 

Functions $\Psi_1$ and $\Psi_2$  form a  
complete set of orthogonal
functions $\{ \Psi_i \}$ which diagonalize the
interaction matrix $T$
\beq \label{GT1}
A^{i'k'}_{i,k}=<\Psi_i\,\Psi_k|\mathbf{T}|\Psi_{i'}\,\Psi_{k'}>=
A_{i,k}\,\delta_{i,i'}\,\delta_{k,k'}.
\eeq
The unitarity constraints have  the form
\beq \label{UNIT}
2\,\mbox{Im}\,A_{i,k}\left(s,b\right)=|A_{i,k}\left(s,b\right)|^2
+G^{in}_{i,k}(s,b),
\eeq
where $G^{in}_{i,k}$ denotes the contribution of all non 
diffractive inelastic processes,
i.e. it is the summed probability for these final states to be
produced in the scattering of a state $i$ off a state $k$. In \eq{UNIT} 
$\sqrt{s}=W$ denotes the energy of the colliding hadrons, and $b$ 
the impact  parameter.
A  solution to \eq{UNIT} at high energies, has the eikonal form 
with an arbitrary opacity $\Omega_{ik}$, where the real 
part of the amplitude is much smaller than the imaginary part.
\beq \label{A}
A_{i,k}(s,b)=i \Lb 1 -\exp\Lb - \Omega_{i,k}(s,b)\Rb\Rb,~~~~~~~
G^{in}_{i,k}(s,b)=1-\exp\Lb - 2\,\Omega_{i,k}(s,b)\Rb.
\eeq
\eq{A} implies that $P^S_{i,k}=\exp \Lb - 2\,\Omega_{i,k}(s,b) \Rb$, is 
the probability that the initial projectiles
$(i,k)$  reach the final state interaction unchanged, regardless of 
the initial state re-scatterings.
\par

 The first approach is to use the eikonal approximation for $\Omega$ in which
 \beq \label{EAPR}
 \Omega^{\pom}_{i,k}(r_\bot, Y - Y_0,b)\,\,=\,\int d^2 b'\,d^2 b''\,
 g_i\Lb \vec{b}',m_i\Rb \,G^{\mbox{\tiny dressed}}\Lb G_\pom\Lb r_\bot,
 Y - Y_0, \vec{b}''\Rb\Rb\,g_k\Lb \vec{b} - \vec{b}'\ - \vec{b}'',m_k\Rb 
 \eeq 
 where $m_i$ denote the masses, which is introduced phenomenologically to
 determine the $b$ dependence of $g_i$ (see below).
However, we do not have any reason to trust the eikonal approximation, 
which we  discovered,  is not sufficient to fit the experimental data.  


  {\it The model: phenomenological feedback for the theoretical approach.}


 We propose a more general approach, which takes into account the new
 small parameters, that are determined by fitting to the experimental 
data
 (see Table 1 and \fig{amp} for notation):
 \beq \label{NEWSP}
 G_{3\pom}\Big{/} g_i(b = 0 )\,\ll\,\,1;~~~~~~~~ m\,\gg\, m_1 
~\mbox{and}~m_2
 \eeq
 
The first equation means that we can develop the approach for the
 Pomeron interactions in which only terms that are  proportional to
 $g_i G_\pom G_{3 \pom}$
 are taken into account, while the contributions of the order of 
 $G_{3 \pom}  G_\pom G_{3 \pom} $ are   negligibly small. 
Therefore, using the first small parameter of \eq{NEWSP}, we  see 
 that the main contribution stems from the net diagrams
 shown in \fig{amp}-b\cite{GLMNET}.

 The second equation in \eq{NEWSP} leads to the fact that $b''$ in 
\eq{EAPR} is much
 smaller than $b$ and $ b'$,
  therefore, \eq{EAPR} can be re-written in
 a simpler form
 \beq \label{EAPR1}
 \Omega^{\pom}_{i,k}(r_\bot, Y - Y_0, b)\,=\,\underbrace{\Bigg(\int d^2 b''\,
G^{\mbox{\tiny dressed}}\Lb
 G_\pom\Lb r_\bot, Y - Y_0, \vec{b}''\Rb\Rb\Bigg)}_{\tilde{G}^{\mbox{\tiny dressed}}\Lb r_\bot, Y - Y_0\Rb}\,\int d^2 b' g_i\Lb
 \vec{b}'\Rb \,g_k\Lb
 \vec{b} - \vec{b}'\Rb 
\eeq

 We can see  that the proton-proton interaction  is similar to
 the nucleus-nucleus interaction. For a  nucleus interaction
 $g_i \propto A^{1/3} \,\gg\,G_{3P}$ and $R_A \gg R_N$, where
 $R_A$ and $R_N$ denote the nucleus and nucleon radii, respectively.
 \eq{NEWSP} shows the same  hierarchy of the vertices and radii
 is present in proton proton scattering.  
 The sum of these diagrams\cite{GLM2CH} leads to the following expression 
for $
 \Omega^{\pom}_{i,k}(s,b)$
 \beq \label{OMEGA}
\Omega^{\pom}_{i, k}\Lb r,  Y-Y_0; b\Rb~=~ \int d^2 b'\,
\,\,\,\frac{ g_i\Lb\vec {b}'\Rb\,g_k\Lb\vec{b} -
 \vec{b}'\Rb\,\tilde{G}^{\mbox{\tiny dressed}}\Lb r, Y - Y_0\Rb
}
{1\,+\,G_{3\pom}\,\tilde{G}^{\mbox{\tiny dressed}}\Lb r, Y - Y_0\Rb\left[
g_i\Lb\vec{b}'\Rb + g_k\Lb\vec{b} - \vec{b}'\Rb\right]} ;~~~
g_i\Lb b \Rb\,=\,g_i \,S_p\Lb b; m_i \Rb ;
\eeq
where
\beq \label{SB}
S_p\Lb b,m_i\Rb\,=\,\frac{1}{4 \pi} m^3_i \,b \,K_1\Lb m_i b
 \Rb~~~\xrightarrow{\mbox{Fourier image} }~~~\frac{1}{\Lb 1
 + Q^2_T/m^2_i\Rb^2};~~
\tilde{G}^{\mbox{\tiny dressed}}\Lb r, Y -Y_0\Rb\,\\\
=\,\,\int d^2 b
 \,\,G^{\mbox{\tiny dressed}}\Lb r, Y-Y_0, b\Rb
 \eeq

 Formula of \eq{OMEGA} describes the net diagrams where the 
Green function of the BFKL Pomeron is replaced by the dressed
 Pomeron Green function, as  is shown in \fig{amp}-b.


The impact parameter  dependence  of $S_p\Lb b,m_i\Rb$ is purely 
phenomenological,
  \eq{SB} which has a form of the electromagnetic proton  form 
factor,
 leads to the correct ($\exp\Lb - \mu b\Rb$) behavior at large 
$b$\cite{FROI},
 and has the correct behavior at large $Q_T$, which has been calculated in 
the
 framework of perturbative QCD \cite{BRLE}. We wish to draw the reader's 
attention to the fact 
 that $m_1$ and $m_2$ are the two dimensional scales in a hadron, which in
 the framework of the constituent quark model,  we assign to the size of 
the
 hadron ($R_h \propto 1/m_1$), and the size of the constituent quark
 ($R_Q 
\propto 1/m_2$).
Note  that  $\tilde{G}^{\mbox{\tiny dressed}}\Lb Y - Y_0\Rb$ does not 
depend
 on $b$.  In all previous formulae, the value of the triple BFKL Pomeron
 vertex
 is known: $G_{3 \pom} = 1.29\,GeV^{-1}$.
  
\begin{table}[h]
\begin{tabular}{|l|l|l|l|l|l|l|l|l|}
\hline
 &$\lambda $ & $\phi_0$ ($GeV^{-2}$) &$g_1$ ($GeV^{-1}$)&$g_2$
 ($GeV^{-1}$)& $m(GeV)$ &$m_1(GeV)$& $m_2(GeV)$ & $\beta$ \\
\hline
Pomeron & 0.38 & 0.0019 & 110.2 & 11.2 & 5.25 & 0.92& 1.9& 0.58\\
\hline
 &$\alpha_{\pom'}(0) $ &$ \alpha_{\omega}(0) $ &$g^{\pom'}_1$ ($GeV^{-1}$)&$g^{\pom'}_2$
 ($GeV^{-1}$)& $R^{\pom'}_{0,1}(GeV^{-1})$ &$R^{\pom'}_{0,2}(GeV^{-1})$& $g^{\omega}_1=g^{\omega}_2 $( $GeV^{-1}$)& $ R^{\omega}_{0,1} =R^{\omega}_{0,2}$
 ($GeV^{-1}$) \\
 \hline
Reggeons& 0.55& 0.55 & 2.937&  5.365 & 2.18&8.633& 3.61 &  2.611 \\
\hline 
  \end{tabular}
\caption{Fitted parameters of the model. The parameters for the
 Pomeron channel are taken from  Ref.\protect\cite{GLM2CH,GLP}.  Reggeons
 parameters  were derived from a fit  to the data at low energies
 (see \fig{fit}).    The value of $\alpha_{\pom'}(0)\,=
\,\alpha_{\omega}(0)$ is fixed
 from our fit for the DIS structure function $F_2$ \cite{GLP}.
  $\alpha'_{\reg}$ is taken to be equal to $1 \,GeV^{-2}$ which
 comes from the reggeon trajectories in the resonance region.
  $\chi^2/d.o.f.$ \,=\,1.2.}
\label{t1}
\end{table}.

{\it Theoretical background: the secondary Reggeons.} \\

Unfortunately, perturbative QCD cannot lead to an understanding of
 the nature of the secondary Regge poles, which describe the energy
 behavior of quasi-elastic processes with non vacuum quantum numbers
 in t-channel. We have abundant experimental confirmations of these
 contributions, as well as  acumen, that the existence of  
secondary Reggeons is ultimately related to the production of a rich 
variety
 of resonances. Therefore, the secondary Reggeons remain an open question, 
that has 
to be solved in non-perturbative QCD. At the moment we  assume
 the pure phenomenological approach for the contribution of the secondary 
Reggeons, replacing $\Omega^{\pom}$ by the sum: $\Omega^{\pom}\,+
\,\Omega^{\reg}$, where $\Omega^{\reg}$ is equal to
\beq \label{REG}
\displaystyle{\Omega^{\reg}_{i,k}\,\,=\,\,\Omega^{\pom'}_{i,k}\,\,\pm\,
\,\Omega^{\omega}_{i,k}~~~\mbox{with}~~~\Omega^{\pom'(\omega)}_{i,k}\,
\,=\,\,\frac{g^{\pom'(\omega)}_{i}\,g^{\pom'(\omega)}_{k}}{
\pi\,R^2_{i,k}\Lb Y\Rb}\,e^{ -
 \frac{b^2}{R^2_{i,k}\Lb Y\Rb}}\,\Lb \frac{s}{s_0}\Rb^{\alpha_{\pom'
(\omega)}\Lb 0\Rb}}
\eeq
Sign $+$ ($-$) in \eq{REG} relates to proton-antiproton (proton-proton)
 scattering.
\beq \label{RIK}
R^2_{i,k}\Lb Y\Rb\,\,=\,\,R^2_{0, i}\,\,+\,\,\,R^2_{0,
 k}\,\,+\,\,\alpha'_{\reg}\,Y
\eeq

 We chose the intercepts of $ \alpha_{\pom'}\Lb 0\Rb\ $ and 
$ \alpha_{\omega}\Lb 0\Rb\ $
 to be equal, in the spirit of the duality between resonances and 
Regge exchanges that leads to signature degeneracy.

 All parameters were determined
 from a fit of the relevant experimental data. In our attempts to describe 
DIS data \cite{GLP0} 
we fixed $\alpha_{\pom'}\Lb 0\Rb\,\,=\,\,0.55$. It should be stressed
 that \eq{REG}  is written for the imaginary part of the amplitude.

 We determine  the real part of the amplitude, 
using  dispersion  relations.


{\it Physical observables:}


In this paper we concentrate our efforts on  the description of the total,
 elastic cross sections and the elastic slope in the region of
 low energies starting from $W = 30\,GeV$.

 For the completeness of presentation, we give the expressions
 for the physical observables that we used in this paper. They can be
 written as follows
\bea
\mbox{elastic~~ amplitude}:&~~&a_{el}(s, b )\,=\,i \Lb \alpha^4 A_{1,1}\,+\,2 \alpha^2\,\beta^2\,A_{1,2}\,
+\,\beta^4 A_{2,2}\Rb; \label{OBS}\\
\mbox{elastic cross section}:&~~~&\sigma_{tot}\Lb s \Rb\,=\,2 \int d^2 b\, a_{el}\Lb s, b\Rb;~~~\sigma_{el}\Lb s \Rb\,=\, \int d^2 b \,|a_{el}\Lb s, b\Rb|^2;\nn\\
\mbox{elastic slope}:&~~~& B_{el}\Lb s \Rb\,\,=\,\,\h \int d^2 b\, b^2\,a_{el}\Lb s, b\Rb\Bigg{/}\int d^2 b\, a_{el}\Lb s, b\Rb.\nn
\eea

As has been mentioned, we wish to study the energy behavior of the 
parameter
 $\rho = \mbox{Re/Im}$. To find the real part of the amplitude we use the
 dispersion relation as  suggested in Ref.\cite{BKS} (see also
 Ref.\cite{REIMDR}).
For prorton-proton scattering the expression for $\rho$ has the form:

\beq \label{PORHO}
\rho\,\,=\,\,\frac{1}{\sigma^{pp}_{tot}}  \frac{\pi}{4}\left\{
 \frac{d }{d \ln\Lb s/s_0\Rb}\Big( \sigma^{pp}_{tot} +
 \sigma^{\bar{p}p}_{tot} \Big)\,\,+\,\frac{1}{s} \,\frac{d}{d
 \ln\Lb s/s_0\Rb}\Big(s\Lb  \sigma^{pp}_{tot} - \sigma^{\bar{p}p}_{tot}
 \Rb\Big)\right\}
\eeq
 In \eq{PORHO} we replace $\tan\Lb \h \pi \frac{d}{ d \ln s}\Rb$ by $
 \h \pi \frac{d}{ d \ln s}$ since  both $ \frac{d }{d \ln\Lb s/s_0\Rb}
\Big( \sigma^{pp}_{tot} + \sigma^{\bar{p}p}_{tot} \Big)$ and $\frac{1}{s}
 \,\frac{d}{d \ln\Lb s/s_0\Rb}\Big(s\Lb  \sigma^{pp}_{tot} - 
\sigma^{\bar{p}p}_{tot} \Rb\Big)$  turn out to be small.
 
 From \eq{A}, \eq{OBS}  and \eq{PORHO} one can see that $\rho\,\propto 
 ~\int d^2 b\, \frac{d \Omega_{i,k}(s,b)}{ d \ln s} \exp\Lb - 
\Omega_{i,k}(s,b) \Rb\,\,\xrightarrow{ s \gg s_0} \,0$. This
 equation shows that $\rho$ is small and decreases at high energies. 
  The values of $\rho$ crucially depend on the amount of the shadowing 
  which we incorporated in the model. As we see below, our model, based on
 the CGC approach, provides the degree of  shadowing which is in accord 
with the
 experimental data. 
 On the other hand, the small value and the behaviour of $\rho$ provides
  hope that we can find the odderon contribution in $\rho$ at high energies.

 { \it Results of the fit.}

  The parameters, related to the Pomeron interaction 
(see first row in Table 1), have been determined in our previous
 papers (see Refs.\cite{GLP,GLP0}) by fitting to  very
 high energies W = 0.574\, $\div$ \,13 \,TeV data. In this paper
 we extract the parameters of the secondary reggeons
   using  data for $\sigma_{tot}$, $\sigma_{el}$
 $B_{el}$  at low energies W =  20 \,$\div$  574 \,GeV.  In
 doing so we pursued two goals:
 to provide independent estimates of the value of the secondary 
reggeon contribution for W \,$\geq$ 0.574\,TeV; and to organize
 the fit in a such way that the abundance of  more precise data
 at lower energies will not spoil the fit at high energies.  
 
   In addition, we fix the
 intercept of the $\pom'$ -Reggeon to the same value we found 
in the
  DIS\cite{GLP} fit. 
 The parameters of the secondary reggeons that we extracted from the fit, 
  are shown in the second row of Table 1. Using these parameters we 
estimate
 that the secondary reggeon contribution for W $\geq$ 0.574 \,TeV is
 negligibly small (<  1\%). It means that the behaviour of $\rho$ at
 these energies is determined only by the Pomeron contributions, assuming
 that there is no  odderon contribution. 
 
       \begin{figure}[ht]
    \centering
  \leavevmode
  \begin{tabular}{c c c}
      \includegraphics[width=6cm]{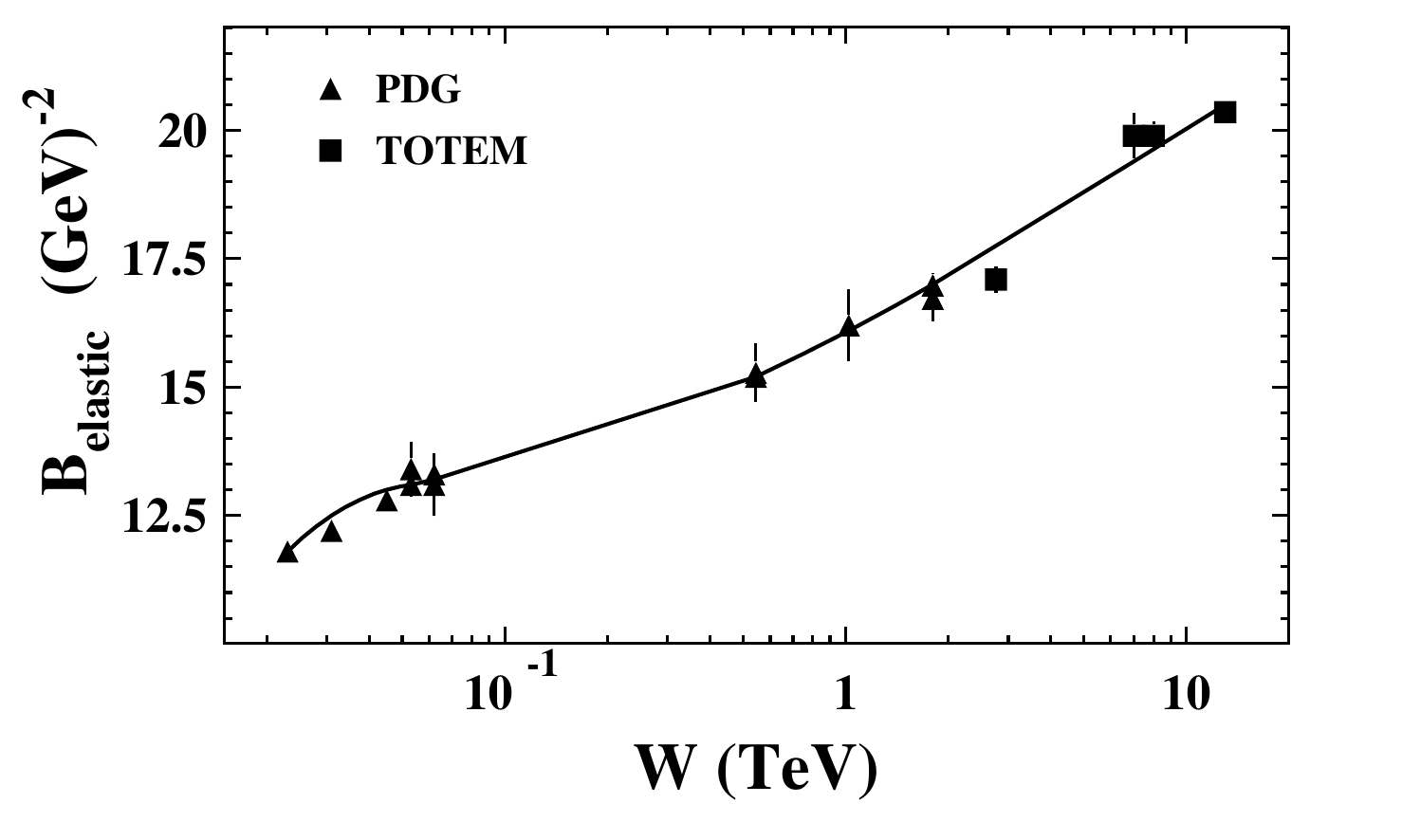}  &\includegraphics[width=6cm]
{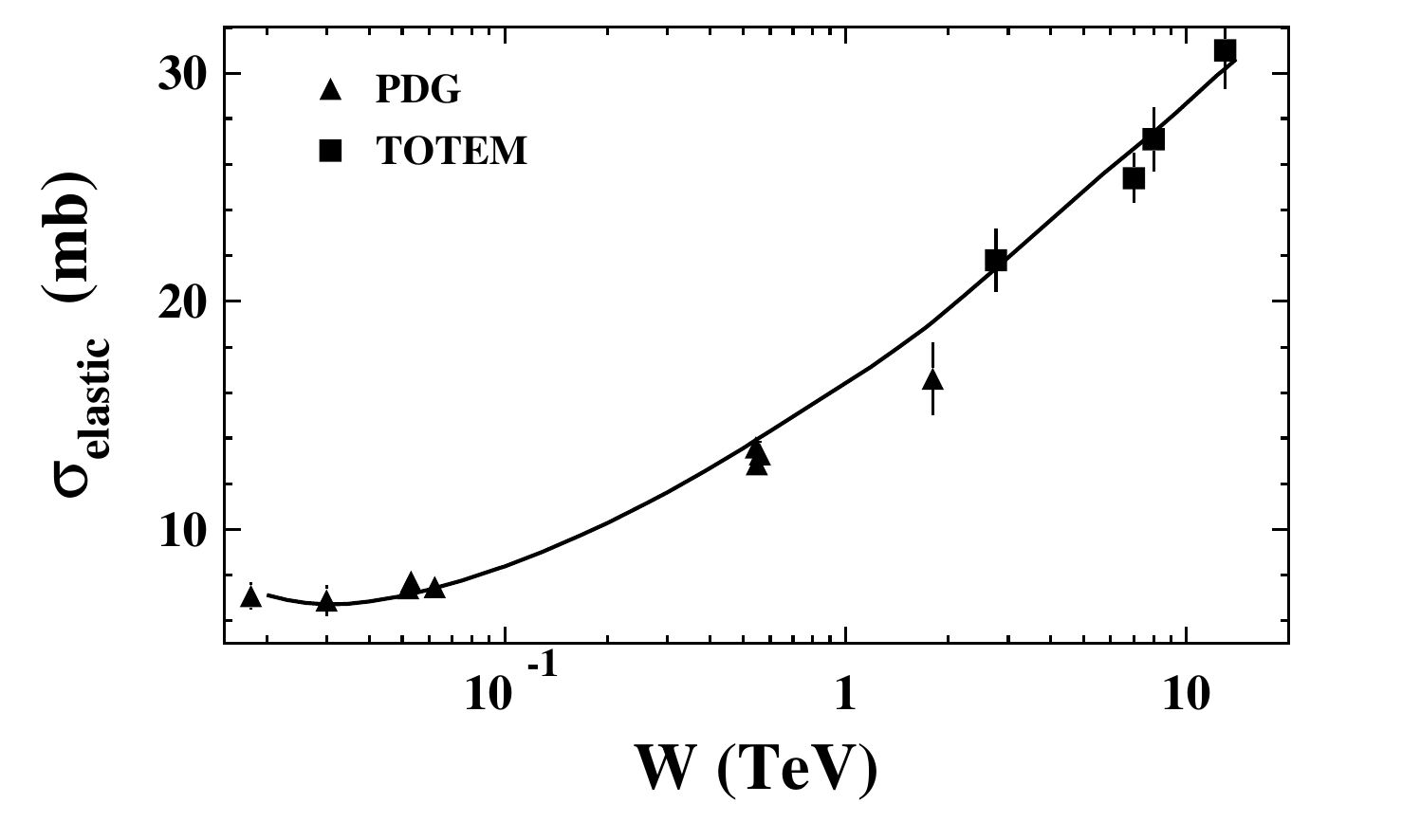}&\includegraphics[width=6cm]{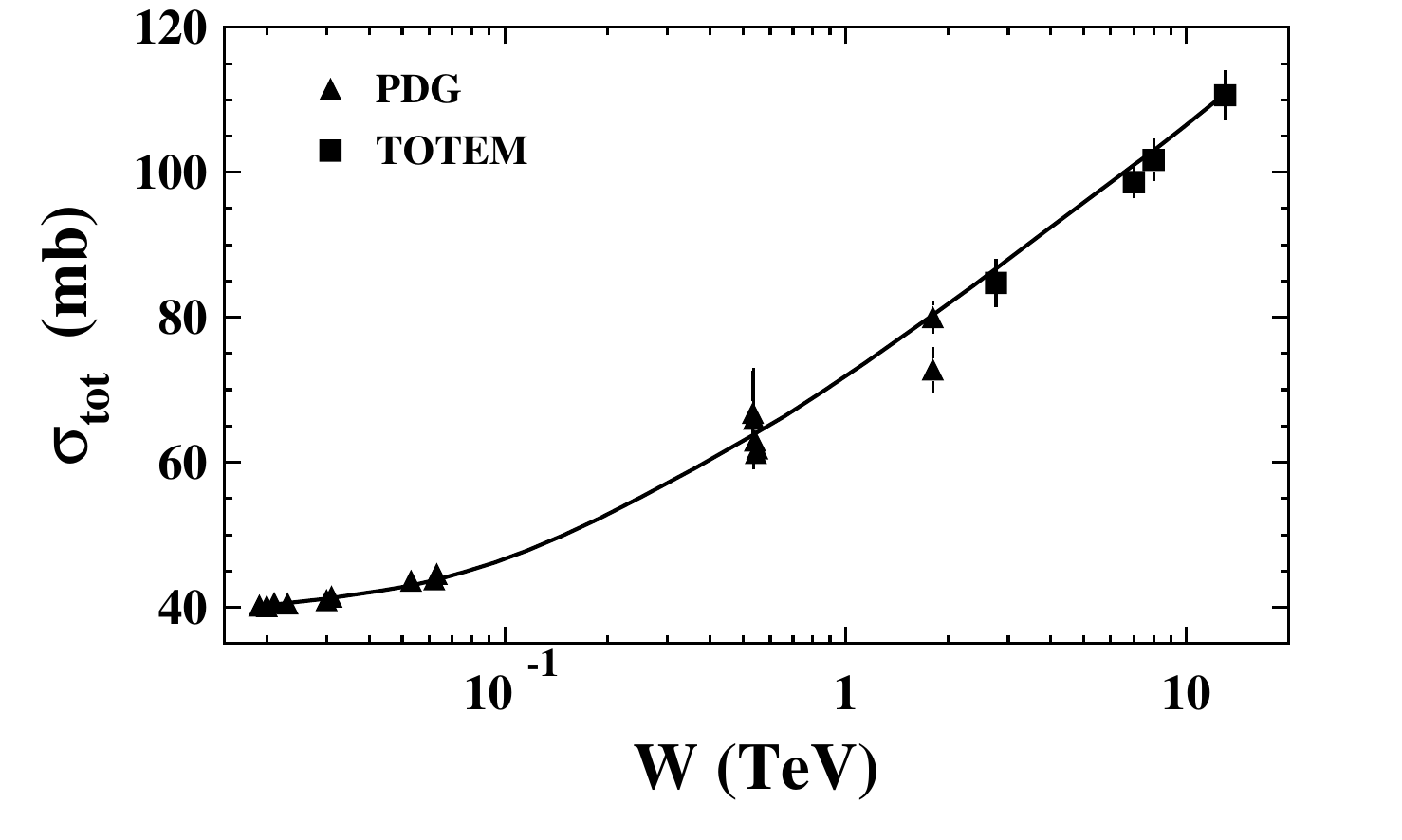} \\
      \fig{fit}-a&\fig{fit}-b& \fig{fit}-c\\
      \end{tabular}
      
 \caption{ The energy behaviour of $\sigma_{tot},\sigma_{el}$ and the
 slope $B_{el}$ for proton-proton scattering in our model. 
 Data are taken from Refs.\cite{PDG,TOTEMRHO}.   }
 \label{fit}
   \end{figure}

\begin{figure}[ht]
    \centering
  \leavevmode
      \includegraphics[width=12cm]{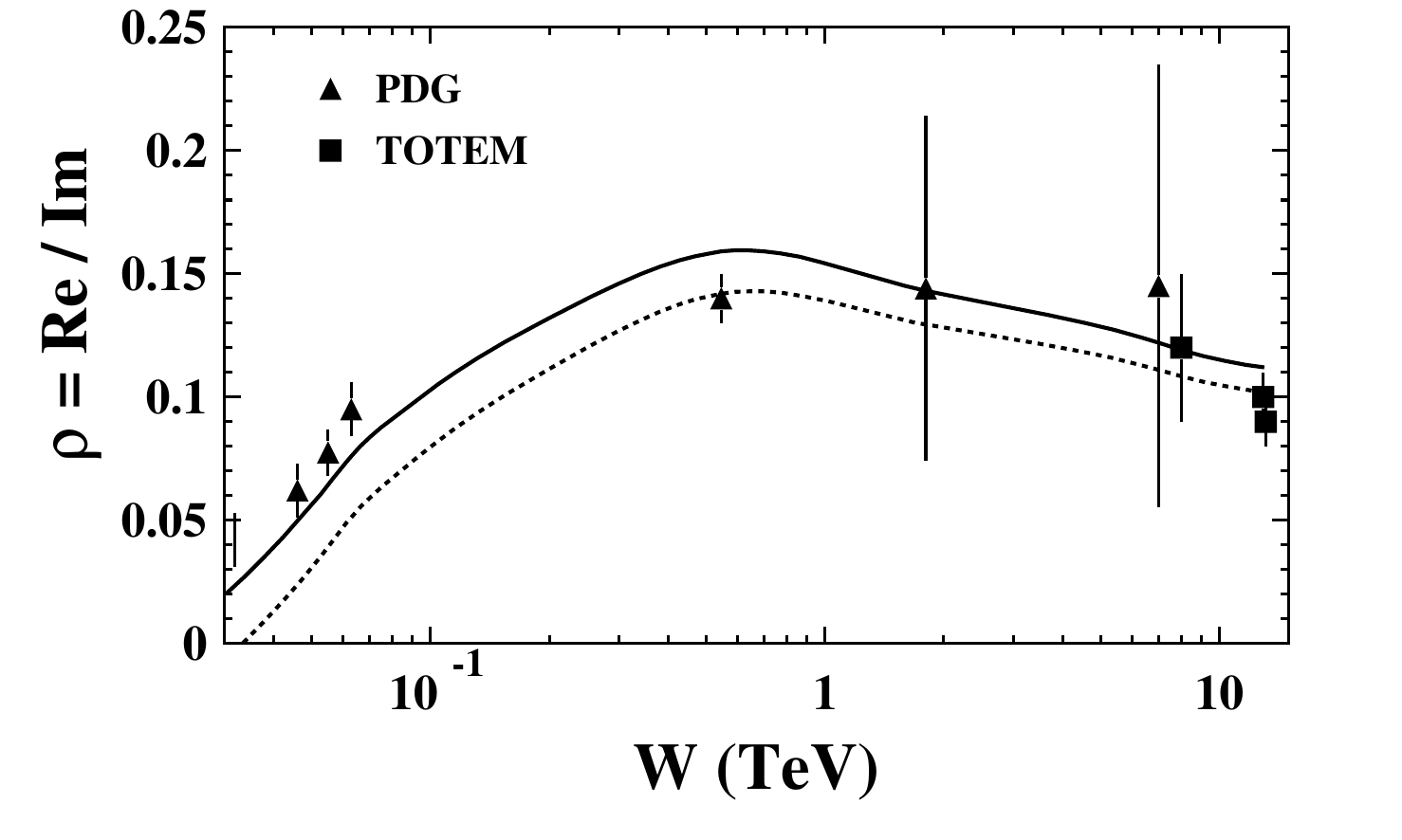}
 \caption{$\rho$ = Re/Im for proton-proton scattering
  versus $W =\sqrt{s}$. Data are taken from PDG \cite{PDG} and
 from the TOTEM papers \cite{TOTEMRHO}. The solid line shows our
 predictions while the dotted one presents the estimates for the
 value of $\rho$, when adding the odderon contribution $1.1 mb$ at W=13 
TeV, 
to our model.}
\label{rho}
   \end{figure}

In \fig{rho} we present our prediction for $\rho$ . 
One can see that even without an additional odderon contribution
 our model predicts that at $W = 13 \,TeV$ $\rho \approx 0.1$. This result 
is in a good agreement with the TOTEM data.
   %

 ~

 {\it Conclusions.}\\

\fig{rho} shows the main result of this paper: our model leads to
 the $\rho = \mbox{Re/Im} \approx 0.1 $ at W=13\,TeV.
It  reproduces the TOTEM data without assuming the additional 
contribution of the odderon. On the other hand the TOTEM data,
 in the framework of our model, allows an odderon contribution
 of the order of 1 mb at W=13  TeV. 
  Since the odderon gives the real contribution, it does not affect
 our fitting procedure. We expect that the odderon contribution  does
 not depend on energy and will lead to smaller values of $\rho$ at low
 energies as  is shown by the dotted line in \fig{rho}.  
  Such a 
contribution does not describe  the data at low energies, in 
spite 
of improving the agreement with the data at high energies, including 
 the TOTEM measurements. 
 
  In perturbative QCD the odderon contribution can only
 depend logarithmically  on energy.
 However, it is shown in Ref.\cite{KS,KOLEB},  using the solution
 of Ref.\cite{LT}, that the shadowing, non-linear
 corrections suppress the energy behavior of the odderon at high 
energies, generating 
the
 survival probability damping factor.  In our model the odderon
 contribution to $\rho$ is screened
 as
 \beq \label{OSC}
 \rho_{\mbox{\tiny{odderon}}}\,\,=\,\,\int d^2 b \,{\rm O}\Lb s, b\Rb \Bigg( 1 -  a_{el}\Lb s,b\Rb\Bigg)\Bigg{/}\int d^2 a_{el}\Lb s,b\Rb
 \eeq
 \eq{OSC} shows that the odderon contribution is suppressed at high
 energies qualitatively in the same way as the Pomeron contribution
 to $\rho$. Therefore, we are of the opinion that 
 $\rho$ is not the appropriate observable to determine the existence of 
the 
odderon. In the model
 we can estimate the contribution of the odderon at lower energies 
taking into account that the impact parameter dependence of the odderon
 which  is expected to be concentrated at smaller $b$ than that for the 
Pomeron\cite{KP}.
 In this case \eq{OSC} takes the form
 
  \beq \label{OSC1}
 \rho_{\mbox{\tiny{odderon}}}\,\,=\,\,\int d^2 b \,{\rm O}\Lb s, b\Rb \Bigg( 1 -  a_{el}\Lb s,b=0\Rb\Bigg)\Bigg{/}\int d^2 a_{el}\Lb s,b\Rb
 \eeq  
 
 Numerically, \eq{OSC1} leads to  $ \int d^2 b \,{\rm O}\Lb s, b\Rb$ = 
 23 mb at W=13\,TeV. Since ${\rm O}\Lb s, b\Rb$ does not depend on energy,
 this large value means that the odderon term should be essential at lower
 energies, and should  be taken in consideration together with the 
secondary
 reggeons.  Indeed, at $W = 21.2\,GeV$  the odderon contribution
 from \eq{OSC1} gives ${\rm Re A\Lb S,t=0\Rb} = 8.8 \,mb$ while the 
contribution of the secondary reggeons to the total cross is equal to
 14.6\,mb.
 Therefore, we do not think that $\rho$ at high energies  is an
 appropriate observable for determining the odderon contribution.

   In this paper we demonstrated that  two effects: the decrease 
with energy as well as the value of $\rho$ ; and the suppression of 
the odderon contribution stem from the amount of the shadowing correction
 that is
 originates from the CGC/saturation approach of our model.
 
 As we have discussed our  model also  describes the wide
 range of the experimental observables.
 We believe that these  successes provide  a strong argument in favour 
of the 
CGC/saturation  approach.

  {\it Acknowledgements.} \\
   We thank our colleagues at Tel Aviv University and UTFSM for
 encouraging discussions.
 This research was supported by the BSF grant   2012124, by 
   Proyecto Basal FB 0821(Chile) ,  Fondecyt (Chile) grants  
 1140842 and 1180118 and by   CONICYT grant PIA ACT1406.

 \end{document}